\documentclass[reprint,amsmath,amssymb,aps,nolongbibliography]{revtex4-2}
\pdfoutput=1
\usepackage{mathtools}
\usepackage{graphicx}% Include figure files
\usepackage{subfig}
\usepackage{dcolumn}% Align table columns on decimal point
\usepackage{bm}% bold math
\usepackage{hyperref}% add hypertext capabilities
\usepackage{xcolor}
\usepackage{appendix}
\usepackage[sfdefault=cmbr]{isomath}
\def \t{\tensorsym}
\def \lb{\left}
\def \rb{\right}

\def \bnabla{\boldsymbol{\nabla}}

\def \bOmega{\mathbf{\Omega}}

\def \para{\parallel}

\def \rhob{\bar{\rho}}
\def \rhoh{\hat{\rho}}
\def \rhot{\tilde{\rho}}

\def \bsigma{\boldsymbol{\sigma}}

\def \bsigmab{\bar{\boldsymbol{\sigma}}}

\def \bzero{\mathbf{0}}

\def \fB{\mathcal{B}}

\def \bD{\mathbf{D}}

\def \bF{\mathbf{F}}

\def \tF{\mathsf{\t F}}

\def \tGb{\mathsf{\t{\bar{G}}}}

\def \bI{\mathbf{I}}

\def \bL{\mathbf{L}}

\def \bQ{\mathbf{Q}}

\def \tR{\mathsf{\t R}}

\def \bS{\mathbf{S}}

\def \tTb{\mathsf{\t{\bar{T}}}}

\def \bU{\mathbf{U}}
\def \bUb{\bar{\mathbf{U}}}

\def \tU{\mathsf{\t U}}
\def \tUb{\mathsf{\t{\bar{U}}}}

\def \fV{\mathcal{V}}

\def \bX{\mathbf{X}}

\def \be{\mathbf{e}}

\def \bf{\mathbf{f}}

\def \bg{\mathbf{g}}
\def \bgh{\hat{\bg}}

\def \bk{\mathbf{k}}

\def \bn{\mathbf{n}}

\def \pt{\tilde{p}}
\def \bp{\mathbf{p}}

\def \rt{\tilde{r}}

\def \br{\mathbf{r}}

\def \brt{\tilde{\br}}

\def \bu{\mathbf{u}}
\def \bub{\bar{\bu}}

\def \but{\tilde{\bu}}

\def \bx{\mathbf{x}}

\begin{document}

\title{\textit{Densitaxis}: Active particle motion in density gradients}

\author{Vaseem A. Shaik}
\author{Gwynn J. Elfring}%
 \email{gelfring@mech.ubc.ca}
\affiliation{Department of Mechanical Engineering, Institute of Applied Mathematics,\\
University of British Columbia, Vancouver, BC, V6T 1Z4, Canada}

\date{\today}

\begin{abstract}
Organisms often swim through density stratified fluids. In this Letter, we investigate the dynamics of small active particles swimming in density gradients and report theoretical evidence of taxis as a result of density stratification (\textit{densitaxis}). Specifically, we calculate the effect of density stratification on the dynamics of a force-free spherical squirmer and show that density stratification induces reorientation that tends to align swimming either parallel or normal to the gradient depending on the swimming gait. In particular, particles that propel by generating thrust in the front (pullers) rotate to swim parallel to gradients and hence display (positive or negative) densitaxis, while particles that propel by generating thrust in the back (pushers) rotate to swim normal to the gradients. This work could be useful to understand the motion of marine organisms in ocean, or be leveraged to sort or organize a suspension of active particles by modulating density gradients.
\end{abstract}

\maketitle

Organisms often swim through inhomogeneous environments experiencing gradients in heat, light, nutrients, temperature or salinity. They frequently respond to these inhomogeneities by reorienting and changing speed, exhibiting a directed motion called \textit{taxis}.  Particle motion in gradients of light \cite{Witman1993, Jekely2008, Jekely2009}, nutrients \cite{Berg1972, Berg2004}, and recently viscosity \cite{Liebchen2018, Datt2019, Stehnach2021, Coppola2021, Shaik2021a, Gong2023}, have been relatively well studied and even exploited to sort or organize active matter systems \cite{Arlt2018, Frangipane2018, Fernandez-Rodriguez2020, Shaik2023}. However, motion through density gradients and any resulting taxis is relatively less well understood despite the prevalence of density stratification in lakes, ponds, and oceans. Density gradients in aquatic bodies are generally caused by corresponding gradients in temperature or salinity and have been known to hinder the vertical motion of organisms like euphausiids and dinoflagellates \cite{Bergstrom1997, Jephson2009}, play a role in the formation of algal blooms \cite{Sherman1998}, and possibly reduce mixing induced by marine organisms \cite{Shaik2020a, Shaik2020}.

The current understanding of (living or non-living) swimmers in density gradients is largely based on the analyses of various model active particles in vertical gradients (see \cite{Ardekani2017, Magnaudet2020, More2023} for a review). Initial studies found the flow due to a point-force or a force-dipole placed in a density-stratified fluid with negligible inertia and weak advection \cite{List1971, Ardekani2010, Wagner2014}. Like in a homogeneous fluid, these singularity solutions represent the far-field flow due to a settling passive particle or a neutrally buoyant active particle in a stratified fluid. Analogous far-field flows with strong advection have also been recently found \cite{Shaik2020, Varanasi2022a}.

Active particles in density gradients not only induce a distinct flow field, but also swim at different speeds compared to homogeneous fluids. Vertically swimming particles have received much attention to address diel vertical migration of marine organisms in search of food. Studies have shown speed changes are indifferent to whether the particle swims up or down the density gradients, for instance, a swimming sheet that propels by passing travelling waves on its surface experiences a speed up or slowdown in density gradients relative to its speed in homogeneous fluids \cite{Dandekar2019}. On the other hand, a squirmer slows down in density gradients if it generates thrust in the back (generally called pushers) but speeds up if generates thrust in the front (pullers), at negligible inertia \cite{Doostmohammadi2012, Shaik2021}. These changes are caused by the buoyancy of the fluid displaced by the particle  \cite{Shaik2020}. Increasing fluid inertia tends to lead to a slowdown for both pushers and pullers \cite{Doostmohammadi2012, More2020}. 

While past research has focused on active particles moving vertically in vertical density gradients, without knowledge of reorientation it is unknown whether active particles would tend to swim in a vertical orientation. Here, we address the question of whether microswimmers tend to align with density gradients from an arbitrary initial orientation as a result of hydrodynamics. Notably, we show that puller-type squirmers rotate to align vertically, displaying positive or negative densitaxis, but pushers rotate to swim horizontally. Implying that density gradients may aid vertical migration for pullers while hinder (or block) it for pushers. We also generalize earlier findings on velocity changes experienced by particles swimming vertically \cite{Shaik2021} to those experienced while swimming in arbitrary direction. 

\begin{figure}[t!]
    \centering
    \includegraphics[scale = 0.45]{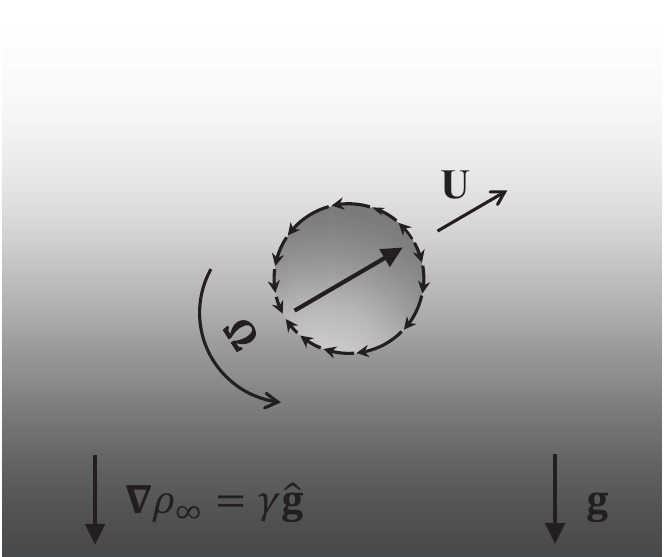}
    \caption{An active particle in a fluid where the density varies linearly in the direction of gravity $\bg$.}
    \label{fig:schematic}
\end{figure}

We consider an active particle moving in an otherwise quiescent Newtonian fluid (see Fig.~\ref{fig:schematic} for a schematic). The fluid density in the absence of particle, $\rho_\infty\left(\bx\right)$, varies spatially due to a similar variation in the temperature or salt concentration. The background (or ambient) density usually varies over a length scale of kilometers in the ocean. At the scale of the small $\mu$m-mm sized particles we consider here, we take the ambient density to vary linearly, $\bnabla\rho_\infty = \gamma \bgh$ where $\gamma>0$ (for stable stratification) is the magnitude of the gradient and $\bgh = \bg/\lb|\bg\rb|$ is a unit vector oriented in the direction of gravity. The quiescent ambient fluid, $\bu_\infty = \bzero$, satisfies a hydrostatic equation $\nabla p_\infty  = \rho_\infty \bg$ where $p_\infty$ is the ambient pressure field.

The introduction of even a passive particle into the fluid will generally disturb the background density, as the properties of the particle, such as thermal conductivity or salt diffusivity, usually differ from those of the fluid. An active particle will moreover generate a flow due to activity that then transports the density. The density field in the presence of the particle is therefore written $\rho\left(\bx\right) = \rho_\infty\left(\bx\right) + \rho'\left(\bx\right)$, where $\rho'$ is the disturbance density. Likewise we may write the pressure field in the presence of the particle as $p=p_\infty+p'$ and because we have assumed an otherwise quiescent fluid, $\bu = \bu'$.

The flow in the presence of the particle is governed by the incompressible Navier-Stokes equations assuming the Boussinesq approximation (which holds for moderate density gradients \cite{Gray1976, Leal2007, Candelier2014}) where the density is taken as constant, $\rhob$, except for in the weight of the fluid. $\overline{\rho}$ is taken as the mean density over a regime containing the particle that is large compared to the particle size. 

The flow field advects temperature or salinity fields whose transport is governed by an advection diffusion equation that ultimately couples with the fluid density. When temperature or salinity variations are weak, gradients in temperature or salinity can be considered linearly proportional to gradients in density and so, in this limit, density transport is also governed by a advection-diffusion equation.

As boundary conditions, we expect that the disturbance due to the particle vanishes far from the particle so that $\bu' \to \bzero$, and $\rho' \to 0$ as $\left|\br\right| \to \infty$, where $\br=\bx-\bX$ with the center of the particle denoted $\bX$. The fluid velocity at the particle surface (denoted $\partial\fB$) satisfies the no slip condition so that $\bu'(\bx\in \partial\fB) = \bU + \bOmega \times \br + \bu^s$, where $\bU$ and $\bOmega$ are the unknown (and to be determined by dynamic conditions) rigid-body translational and rotational velocity of the particle while $\bu^s$ is the velocity on the surface due to activity (deformation or slip). 

We model the active particle as a spherical squirmer of radius $a$. In the squirmer model, the shape of the swimmer is fixed and activity due to any swimming gait is represented simply by a prescribed tangential slip on the particle surface \cite{Lighthill1952, Blake1971, Ishikawa2006}. This model is a reasonable representation of ciliated organisms like \textit{Paramecium} and \textit{Opalina} which propel by synchronously beating numerous cilia on their surface. The slip velocity is typically written as an expansion in Legendre polynomials $\bu^s = \sum_n \frac{2}{n(n+1)}B_nP_n' (\bp\cdot\bn)\bp\cdot(\bI-\bn\bn)$ where $\bp$ is the particle orientation, $\bn$ is the outward unit normal to the particle surface, $P_n$ is the Legendre polynomial of degree $n$, and $P_n'(x)=\frac{d}{dx}P_n(x)$. The polar slip coefficients, $B_n$, are often called `squirming' modes, while the azimuthal slip (or `swirl') is not considered here. In homogeneous Newtonian fluids, only the $B_1$ mode determines the swim speed $U_N = \frac{2}{3}B_1$ (we assume $B_1 \ge 0$ with no loss of generality) whereas the $B_2$ mode gives the slowest decaying flow and hence determines the far-field representation of the particle. The ratio of the first two modes is denoted by $\beta = B_2 / B_1$, with $\beta>0$ denoting pullers and $\beta<0$ pushers. We keep only the first two squirming modes to analyze the effects of density gradients on swimming, as is common practice in other analysis including the swimming under confinement \cite{Li2014, Shaik2017}, in complex fluids \cite{Zhu2011, Zhu2012, Li2014a, Yazdi2014, Yazdi2015, Yazdi2017}, and at finite inertia \cite{Wang2012, Khair2014, Chisholm2016, Li2016}. 

The boundary condition for the density disturbance at the particle surface depends on the scalar responsible for the density variations, we assume the scalar flux into the particle vanishes (impermeable to the salt or insulating to the temperature) which leads to a corresponding no-flux condition for density at the surface of the particle, $\bn\cdot \nabla \rho = 0$ for $\bx\in \partial\fB$.

These equations are closed by conservation of linear and angular momentum of the particle, which determine the unknown rigid-body translational and rotational velocity, $\bU$ and $\bOmega$ respectively.

Generally one expects the density of organisms to be similar to their suspending fluid and for simplicity, we assume here that the particle density exactly matches the background fluid density, $\rho_p = \rho_\infty$ \cite{Lee2019, Shaik2021, Varanasi2022}. As the particle swims, it would hence need to alter its density according to the background density to stay density matched. Such density regulation can be achieved by employing a swim bladder or mechanisms like the carbohydrate ballasting \cite{Villareal2003} or the ion replacement \cite{Boyd2002, Sartoris2010}.

The relevant characteristic scales of the problem are the particle size $a$, the density change across the particle $\gamma a$, the swimming speed in a homogeneous fluid $U_N$, and the viscous pressure scale $\rhob\nu U_N/a$ where $\nu$ is the kinematic viscosity of the fluid. Using these, the governing equations are characterized by three dimensionless numbers: the Reynolds number, $Re = aU_N/\nu$, measures the relative importance of inertia to the viscous forces, the viscous Richardson number, $Ri_v = a^3 N^2/\nu U_N$, measures the importance of buoyancy to the viscous forces, and the P\'eclet number, $Pe = a U_N/\kappa$, measures the importance of the advective to the diffusive transport rate of the density with diffusivity $\kappa$. The Prandtl number $Pr$ is the ratio of the momentum to the mass diffusivity $\nu/\kappa$ and it can be written in terms of $Re$ and $Pe$ as $Pr = Pe/Re$. The Brunt-V\"{a}is\"{a}l\"{a} (or buoyancy) frequency, $N = \sqrt{g \gamma/\rhob}$, appearing in $Ri_v$ is the rate at which a vertically displaced inviscid fluid element oscillates in density gradients. For typical values of these dimensionless numbers we consider planktonic organisms and ocean water. \textit{E. coli}, \textit{Paramecium}, and \textit{Daphina} are of size 1-10 $\mu$m, 100 $\mu$m, and 5 mm, and they swim in water at speeds 10 $\mu$m/s, 1 mm/s, and 1 cm/s, respectively \cite{Guasto2012, Noss2012}. Hence the particle size $a$ ranges from 1 $\mu{\rm{m}}-5$ mm while the particle speed $U_N$ varies from 10 $\mu {\rm{m/s}}-1$ cm/s. On the other hand, the viscosity of water $\nu \approx 10^{-6} \, {\rm{m}}^2$/s while $Pr$ ranges from $7$ to $700$ depending on whether variations in density are due to temperature or salinity, respectively. In oceans, lakes and ponds $N$ varies from $10^{-4}-0.3 \,{\rm{s}}^{-1}$ \cite{Thorpe2005}. These attributes can be used to estimate values of the Reynolds number, $Re = 10^{-5}-50$, the viscous Richardson number, $Ri_v = 10^{-15}-1$ and the P\'eclet number, $Pe = 10^{-4}-10^4$. 

Here, we focus on the regime where inertia is negligible, the limit $Re \rightarrow 0$, with weak but non-negligible effects of buoyancy, and the advection of density (specified more precisely later). This parameter regime corresponds to the motion of small organisms ($\sim 10 - 1000 \,\,\mu$m) in water stratified with weak gradients. We also assume that the density field is steady in a frame of reference moving with the particle. With these assumptions our governing equations for the fluid become in \textit{dimensionless} form
\begin{gather}
    \nabla \cdot \bu' = 0,
    \label{eqn:cont_final_main}\\
    -\bnabla p' +\nabla^2\bu' = Ri_v \, \rho' \bgh,
    \label{eqn:NS_final_main}\\
     \nabla^2 \rho' = Pe \big((\bu'-\bU)\cdot\bnabla\rho' - \bu' \cdot \bgh\big).
    \label{eqn:Advec-Diff_final_main}
\end{gather}
The diffusivity and the viscosity of the fluid are assumed to be constant to leading order, even though they too can vary spatially due to the variations in temperature or salinity. Furthermore, given that we have assumed the particle density matches that of the fluid, the active particle has no net weight in the fluid and the particle inertia also vanishes, hence the disturbance hydrodynamic force and torque on the particle are zero. Unless otherwise specified equations below are written in dimensionless form.

We use a generalized reciprocal theorem to obtain explicit expressions for the particle velocities $\bU$ and ${\bOmega}$. First we note that the disturbance flow is governed by forced Stokes equations where the forcing is the weight of the fluid relative to background density. The dynamics of a force and torque free active particle immersed in such a Stokes flow is given directly by the formula
\begin{align}
\tU=\bar{\tR}_{\tF\tU}^{-1}\cdot\lb[\tF_s +\tF_f \rb],
\end{align}
where we have used six-dimensional vectors for translational and rotational velocity, $\tU =[\bU, \ \bOmega]$ and force and torque, $\tF =\lb[\bF, \ \bL\rb]$. $\bar{\tR}_{\tF\tU}$ is the resistance tensor for the particle in a homogeneous fluid of viscosity $\rhob\nu$, $\tF_s$ is the thrust generated by an active particle in the same homogeneous fluid and $\tF_f$ accounts for the changes due to the presence of density perturbations. The forces are obtained, using the reciprocal theorem \cite{Elfring2017, Masoud2019}, by projecting onto a known rigid-body motion auxiliary flow. See Appendix \ref{app:reciprocal} for details.

Separating translational and rotational velocities we obtain 
\begin{align}
\bU-\bU_N &= \frac{Ri_v}{8\pi}\bgh\cdot\int_{\partial\fB} \bigg[(\bI+\bn\bn) \int_1^\infty \rho' rdr\nonumber\\
&\quad+(\bI/3 - \bn\bn) \int_1^\infty \frac{\rho'}{r} dr\bigg] dS,\label{eqn:DeltaU_recip_main}\\
\bOmega &= \frac{Ri_v}{8\pi}\bgh\times\int_{\partial\fB} \bn \int_1^\infty \rho'dr dS
\label{eqn:Omega_recip_main}
\end{align}
where we've used the fact that when $Ri_v=0$ the particle moves through a homogeneous Newtonian fluid where the dynamics of a squirmer are well known \cite{Stone1996}, $\bU = \bU_N = \bp$ and $\bOmega = \bzero$. We see that to evaluate the effects of density gradients, we simply need to integrate the disturbance density against a known kernel. It may seem as if we have bypassed the need to solve for the disturbance flow; however the disturbance flow couples to the the disturbance density by way of \eqref{eqn:Advec-Diff_final_main}. 

When advection is weak, density transport \eqref{eqn:Advec-Diff_final_main} is dominated by diffusion near the sphere; however, no matter how small $Pe$ there is always a distance, characterized by the screening length $l_{\rho} \sim 1/Pe$, beyond which advection terms become on the order of diffusion. Likewise for the forced Stokes equations when $Ri_v\ll 1$, no matter how small $Ri_v$, buoyancy forces become comparable to viscous forces at a distance from the particle characterized by the stratification screening length $l_s$ (formally defined below). Solutions depend on the relative order of magnitude of the two parameters, $Pe$ and $Ri_v$ (or $l_s/l_{\rho}$) as there are three regions in the solution.

Scaling arguments can be used to determine that the dominant effect on particle rotation comes from the region far from the particle. See Appendix \ref{app:analysis}. As such, we derive equations valid in the far-field $\left( r \gg 1 \right)$, where appropriate singular forcing terms are added to \eqref{eqn:cont_final_main}-\eqref{eqn:Advec-Diff_final_main} to account for the presence of the force-free active particle to leading order in the far field. Specifically, a force dipole of strength $\bS= 2 \pi \beta \left( 3 \bp\bp - \bI \right)$ and a degenerate quadrupole of strength $\bQ = - 2 \pi \bp$, for the disturbance flow, while the density disturbance generated by the particle has character of concentration-dipole $\bD = -2 \pi \bgh$, coefficients are assumed constant to leading order. We linearize by discarding the nonlinear advection of the disturbance density by the disturbance flow (as the product of small terms), and approximate the particle velocity with its value in homogeneous fluid (which holds for weak density gradients $Ri_v \ll 1$). We solve the singularly forced linearized governing equations in Fourier space to find the Fourier transform of density for arbitrary values of $Pe$ and $Ri_v$; however, we calculate the inverse transform, and hence velocities, only in the asymptotic regimes $l_s\ll l_\rho$, and $l_s\gg l_\rho$. See Appendix \ref{app:analysis} for details.

In the regime $l_s \ll l_{\rho}$, far from the particle diffusion dominates advection. The stratification screening length is found by balancing the buoyancy forces with the viscous forces at $r \sim l_s$ to find $l_s \sim \left( Ri_v Pe \right)^{-1/4}$. We expand the density in Fourier space assuming $(Ri_v Pe )^{1/4}\ll 1$ $(l_s\gg 1)$, and $Pe^{3/4}/Ri_v^{1/4}\ll 1$ $(l_s/l_\rho \ll 1)$, then calculate the angular velocity \eqref{eqn:Omega_recip_main} in real space using the convolution theorem to obtain
\begin{align}
    \bOmega = c\beta \left( Ri_v Pe \right)^{3/4} (\bp\cdot \bgh) (\bp\times\bgh),
     %\label{eqn:Omega-result1}
\end{align}
where $c =  E_E\left(1/\sqrt{2}\right)/4 - E_K \left(1/\sqrt{2}\right)/8\approx 0.1059$ and $E_K$ and $E_E$ are the complete elliptic integrals of the first and second kind. Velocity changes are given in Appendix \ref{app:analysis}.

In the other regime, $l_s \gg l_{\rho}$. Far from the particle now advection dominates diffusion. To determine the stratification screening length we again balance viscous forces and buoyancy forces to obtain $l_s \sim Ri_v^{-1/3}$. We solve for the leading order density for $Ri_v^{1/3}\ll 1$ ($l_s \gg 1$) and $Ri_v^{1/3}/Pe\ll 1$ ($l_s/l_\rho \gg 1$) in Fourier space then numerically inverse transform to obtain the angular velocity ${\bOmega}$ in real space and find that data fits 
\begin{equation}
{\bOmega} =  c\beta  Ri_v \left( \bp\cdot \hat\bg \right) \left( \bp\times\bgh \right),
%\label{eqn:Omega-highPe}
\end{equation}
where the constant of proportionality, $c \approx 0.09375$.

We see that both asymptotic regimes have the same functional form to leading order, writing $\dot{\bp} = \bOmega\times\bp$ we have
\begin{align}
\dot{\bp} =  c\beta \varepsilon (\bp\cdot \bgh) (\bI-\bp\bp)\cdot\bgh,
\end{align}
where only the constant of proportionality $c$ (which is $\approx 10^{-1}$ in both cases) and $\varepsilon = \{ (Ri_v Pe)^{3/4}, Ri_v\}$ differ to leading order depending on the regime. 
In both cases we see that fixed points in the dynamics are when $\bp \para \bgh$ and $\bp \perp \bgh$ with stability depending on the sign of $\beta$. In general, for pullers the stable orientation occurs when $\bp \para \bgh$ whereas for pushers when $\bp \perp \bgh$. In other words, in the presence of density gradients pullers will be reoriented to swim up or down the gradient (depending on whether the initial orientation is up or down), whereas pushers will be reoriented orthogonal to gradients. This is reminiscent of the far-field interaction of pushers and pullers with solid walls \cite{Burke2008}. In that case, the functional form is identical except the wall normal replaces the direction of gravity such that pushers orient parallel to the wall while pullers orient normal and the strength of the interaction decays with the distance from the wall. 

\begin{figure}[t!]
    \centering
    \includegraphics[scale = 0.295]{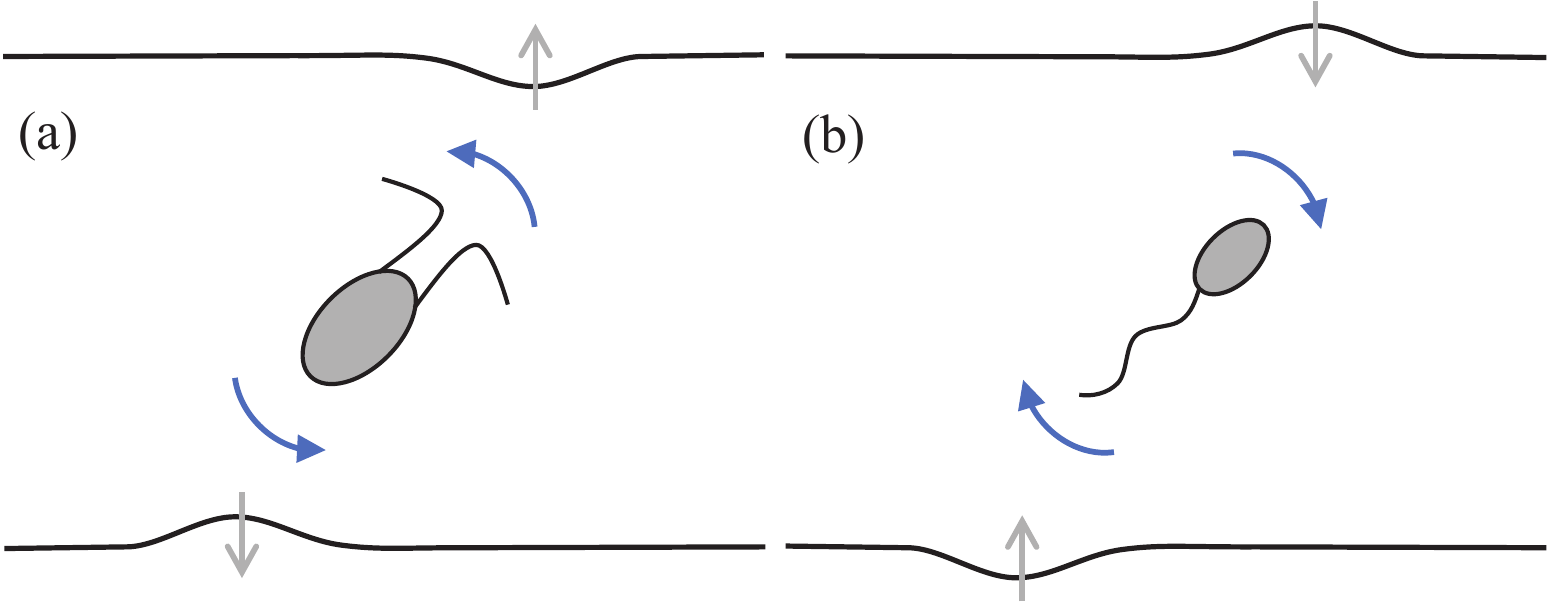} 
    \caption{Active particles deform isopycnals (black lines) differently for (a) pullers and (b) pushers. The resulting flow (baroclinic vorticity) induces reorientation.}
    \label{fig:fig2}
\end{figure}

The mechanism for this rotation is conceptually similar to the far-field interaction of a swimmer with solid walls. The rotation can be understood by analyzing the interaction of far-field dipole flow and density isosurfaces (isopycnals), which we illustrate in a schematic in Fig.~\ref{fig:fig2}. Puller swimmers pull the fluid along their axis toward their body, ultimately perturbing the isopycnals from the horizontal in a manner shown in Fig. 2a. The tendency of isopycnals is to resist deformation and return to their stable state (gray arrows), this induces a flow (specifically baroclinic vorticity shown by blue curved arrows) that rotates the particle. Pushers do opposite to pullers, in that they push fluid along their axis and perturb isopycnals in an opposite manner, hence the induced baroclinic vorticity rotates them in the opposite direction (see Fig. 2b). Swimmers also drive fluid normal to their axis which perturbs the isopycnals differently, but this perturbation and the ensuing baroclinic vorticity is weak and hence not shown here for clarity.

This rotation occurs over a timescale $\tau \sim a/U_N c \left| \beta \right| \varepsilon$. If we assume the particles swim at roughly a body length per second, and $\beta = O(1)$ \cite{Gong2023}, we obtain $\tau \sim 10\, \text{s}/\varepsilon $ depending on the value of $\varepsilon$ (though we have assumed $\varepsilon^{1/3} \ll 1$). In a given density gradient (fixed $\gamma$ or $N$), larger particles gives larger $\varepsilon$ and hence rotate faster. This densitactic rotation is unaffected by thermal fluctuations provided it occurs faster than that due to rotational diffusion $\tau_R^{-1} \equiv D_R = k_B T/8\pi \bar{\rho} \nu a^3$, i.e., $\tau / \tau_R \ll 1$, where $k_B$ is the Boltzmann constant and $T$ is the absolute temperature. The influence of fluctuations decreases with increasing particle size and matters for particles smaller than $O(10\mu m)$ in density gradients relevant to oceanic settings $N\sim 10^{-4} - 0.3\,\text{s}^{-1}$

The densitaxis shown here could have several implications. It could aid or hinder the diel vertical migration of planktonic organisms in ocean depending on their gait and orientation.  It also determines the distribution of a dilute suspension of active particles in density gradients and hence the mixing induced by them. Pullers tend to a nematic state pointing upwards and downwards with equal probability, while pushers tend to swimming in horizontal planes (with isotropic in-plane orientation). These distributions may be important to incorporate into previous models of mixing by microswimmers that assumed an isotropic distribution \cite{Wagner2014}. A further research avenue to investigate would be densitaxis for larger aquatic organisms with finite inertia. Finally, densitaxis might also be leveraged as a mechanism to separate a mixtures of pushers and pullers by imposing density gradients by spatially varying salinity or temperature.

%\acknowledgements
The authors gratefully acknowledge funding (RGPIN-2020-04850) from the Natural Sciences and Engineering Research Council of Canada (NSERC).

\appendix

%Here we provide a detailed mathematical model for the locomotion of active particles (squirmers) in density gradients in Appendix ~\ref{app:model}. We then derive via the reciprocal theorem the general formulas for reorientation and speed changes experienced by the particle in Appendix ~\ref{app:reciprocal} and evaluate these formulas in the asymptotic limits of weak density advection and stratification in Appendix ~\ref{app:analysis}.

\section{\label{app:model}Model}
%\begin{figure}[t!]
%    \centering
%    \includegraphics[scale = 0.5]{fig1.eps}
%    \caption{A schematic showing an active particle moving through density gradients. The particle is of radius $a$ and it translates with velocity $\bU$ and rotates with $\bOmega$. The background density $\rho_\infty$, on the other hand, decreases linearly along the gravity $\bg$. Here $\bgh = \bg/\left| \bg \right|$.}
%    \label{fig:schematic}
%\end{figure}
Here we provide a detailed mathematical model for the locomotion of active particles (squirmers) in density gradients. We consider an active particle moving in an otherwise quiescent Newtonian fluid. The fluid density in the absence of particle, $\rho_\infty\left(\bx\right)$, varies linearly along gravity due to a similar variation in temperature or salt concentration
\begin{align}
\bnabla\rho_\infty = \gamma \bgh.
\end{align}
Here $\gamma>0$ is the magnitude of the gradient and $\bgh = \bg/\lb|\bg\rb|$ is a unit vector oriented in the direction of gravity. Integrating we write the ambient (or background) density as $\rho_\infty = \overline{\rho} + \gamma\bgh\cdot(\bx-\overline{\bX})$ where $\overline{\rho}$ is the mean density over a regime containing the particle that is large compared to the particle size, the center of the average is denoted $\overline{\bX}$, which in practice we treat as the origin. The (stably stratified) quiescent ambient fluid, $\bu_\infty = \bzero$, satisfies a hydrostatic equation
\begin{align}
\nabla p_\infty  = \rho_\infty \bg,
\end{align}
where $p_\infty$ is the ambient pressure field.

The introduction of particle into the fluid generally disturbs the background density, hence the density field in the presence of particle is
\begin{align}
\rho\left(\bx\right) = \rho_\infty\left(\bx\right) + \rho'\left(\bx\right),
\end{align} 
where $\rho'$ is the disturbance density. Likewise we may write the pressure field in the presence of the particle as $p=p_\infty+p'$.

The flow in the presence of the particle is governed by the incompressible Navier-Stokes equations (in the Boussinesq approximation)
\begin{align}
    \nabla \cdot \bu &= 0, \label{eqn:cont}\\
    \rhob \left( \frac{\partial \bu}{\partial t} + \bu \cdot \nabla \bu \right) &= \rho \bg -\bnabla p + \rhob\nu\nabla^2\bu,\label{eqn:NS}
\end{align}
where $\nu$ is the kinematic viscosity of the fluid. Because we have assumed an otherwise quiescent fluid, $\bu = \bu'$. On the other hand, temperature or salinity transport is governed by an advection-diffusion equation, and for weak variations of these fields, density is linearly proportional to them and hence its transport is also governed by a advection-diffusion equation
\begin{equation}
\frac{\partial \rho}{\partial t} + \bu \cdot \nabla \rho = \kappa \nabla^2 \rho,
\label{eqn:density}
\end{equation}
where $\kappa$ is the diffusivity. Here as part of the Boussinesq approximation, we neglect viscous dissipation contribution to this equation which arises from density variations caused specifically by temperature variations.

It is convenient to work with the disturbance variables \cite{Candelier2014, Mehaddi2018}, using the definition above for density and pressure, in \eqref{eqn:cont}, \eqref{eqn:NS}, \eqref{eqn:density}, we obtain
\begin{gather}
    \nabla \cdot \bu' = 0,\\
\rhob\left( \frac{\partial\bu'}{\partial t} + \bu' \cdot \nabla \bu'\right) = \rho' \bg-\bnabla p'+ \rhob \nu \nabla^2 \bu',\\
\frac{\partial \rho'}{\partial t} + \bu' \cdot \nabla \rho' - \gamma \bu' \cdot \bgh  = \kappa \nabla^2 \rho'.
\label{eqn:density-disturb}
\end{gather}
The density transport equation \eqref{eqn:density-disturb} reveals advection of the disturbed density $\bu'\cdot \nabla \rho'$ as well as the background density $\bu' \cdot \nabla \rho_\infty  = - \gamma  \bu' \cdot \bgh$.

As boundary conditions, we expect that the disturbance due to the particle vanishes far from the particle so that
\begin{equation}
    \bu' \to \bzero, \,\, \rho' \to 0 \,\, {\rm{as}} \,\, r = \left|\br\right|\to \infty,
\end{equation}
where $\br = \bx - \bX$ and $\bX$ denotes the position of the center of the particle. The fluid velocity at the particle satisfies the no slip condition so that
\begin{equation}
    \bu'(\bx\in \partial\fB) = \bU + \bOmega \times \br + \bu^s,
    \label{eqn:BC_activeparticle}
\end{equation}
where $\partial\fB$ denotes the particle surface. Here $\bU$ and $\bOmega$ are the unknown rigid-body translational and rotational velocity of the particle while $\bu^s$ is the slip velocity due to activity given by
\begin{align}
\bu^s &= \sum_n \frac{2}{n(n+1)}P_n' (\bp\cdot\bn)\bp\cdot\lb[B_n(\bI-\bn\bn)+C_n\bI\times\bn\rb]
\end{align}
in the squirmer model \cite{Lighthill1952, Blake1971, Ishikawa2006}. Here $\bp$ is the particle orientation, $\bn$ is the outward unit normal to the particle surface, $P_n$ is the Legendre polynomial of degree $n$, and $P_n'(x)=\frac{d}{dx}P_n(x)$. We do not consider any swirl modes $C_n = 0$ but consider only the first two squirming modes $B_n$ and denote their ratio by $\beta = B_2/B_1$. Also the particle surface is impermeable to salt (or insulating to temperature), hence the density field at the particle satisfies a no-flux condition, $\bn \cdot \nabla \rho = 0$ for $\bx\in \partial\fB$, which in terms of disturbance variables is
\begin{align}
    \bn\cdot \nabla \rho' = -\gamma\bn\cdot\bgh \qquad \text{for} \ \bx\in \partial\fB.
\end{align}

%The boundary condition for the density disturbance at the particle surface depends on the scalar responsible for the density variations. Usually, one assumes the scalar flux into the particle vanishes as the particle's surface is impermeable to the salt (or insulating to the temperature). Given the assumed linear relationship between salt/temperature gradients and density gradients we obtain a corresponding no flux condition for density at the surface of the particle, $\bn \cdot \nabla \rho = 0$ for $\bx\in \partial\fB$, which in terms of disturbance variables is written

These equations are closed by conservation of linear and angular momentum of the particle, which determine the unknown translational and rotational velocity, $\bU$ and $\bOmega$ respectively, namely
\begin{gather}
    m_p\frac{d\bU}{dt} = \int_{\partial\fB}\bn\cdot\bsigma' dS + \int_{V_p}(\rho_p-\rho_\infty)\bg dV, \label{eqn:linmom}\\
    I_p\frac{d\bOmega}{dt} = \int_{\partial\fB}\br\times(\bn\cdot\bsigma') dS, \label{eqn:angmom}
\end{gather}
where $m_p$, $\rho_p$ and $I_p$ are the particle mass, mass density, and moment of inertia. The right-hand side is written in terms of disturbance variables, and $\bsigma' = -p' {\bI}+\rhob\nu \left( \nabla \bu' + \left(\nabla \bu'\right) ^T\right )$ is the disturbance stress. By symmetry there is no net torque on the particle due to the hydrostatic pressure on its surface and we assume similarly for its mass (they are not bottom-heavy for instance). We also assume the particle density exactly matches the fluid density at all times, that is $\rho_p = \rho_\infty$ \cite{Lee2019, Shaik2021, Varanasi2022}. %As the particle swims, it hence alters its density according to the background density to stay density matched. Such density regulation can be achieved by employing a swim bladder or mechanisms like the carbohydrate ballasting \cite{Villareal2003} or the ion replacement \cite{Boyd2002, Sartoris2010}.

We now non-dimensionalize the governing fluid equations with the relevant characteristic scales, the particle radius $a$, the density changes across the particle $\gamma a$, the swimming speed in a homogeneous fluid $U_N = 2B_1/3$, and the viscous pressure or stress scale $\rhob\nu U_N/a$. Using the same notation as the dimensional counterparts, the dimensionless governing equations are
\begin{gather}
    \nabla \cdot \bu' = 0,\\
    -\bnabla p' +\nabla^2\bu' = Re \left(\frac{\partial \bu'}{\partial t}+ \bu' \cdot \nabla \bu'\right) + Ri_v \, \rho' \bgh,\\
    Pe \left(\frac{\partial\rho'}{\partial t}+ \bu' \cdot \nabla \rho' - \bu' \cdot \bgh\right) = \nabla^2 \rho'.
\end{gather}
These equations are characterized by three dimensionless numbers: the Reynolds number $Re = aU_N/\nu$, the viscous Richardson number $Ri_v = a^3 N^2/\nu U_N$, and the P\'eclet number $Pe = a U_N/\kappa$, where the buoyancy frequency $N = \sqrt{g \gamma/\rhob}$. Typical value of these dimensionless numbers for the planktonic organisms swimming in oceans, lakes or ponds is \cite{Guasto2012, Noss2012, Thorpe2005} $Re = 10^{-5}-50\,, Ri_v = 10^{-15}-1$, and $Pe = 10^{-4}-10^4$.

Here, we focus on the regime where inertia is negligible, the limit $Re \rightarrow 0$, with weak but non-negligible effects of buoyancy, and the advection of density (a more precise bound on these numbers is provided later). We also assume that the density field is steady in a frame of reference moving with the particle such that 
\begin{align}
\frac{D\rho'}{D t} = (\bu'-\bU)\cdot\bnabla\rho'.
\end{align}
Here we neglect any effect of particle rotation in this equation, as rotation is solely caused by density gradients and hence is small for weak density gradients of interest to us, $\bOmega \sim Ri_v^n \ll \bU$ for $Ri_v \ll 1, n>0$. With these assumptions our governing equations for the fluid become
\begin{gather}
    \nabla \cdot \bu' = 0,
    \label{eqn:cont_final}\\
    -\bnabla p' +\nabla^2\bu' = Ri_v \, \rho' \bgh,
    \label{eqn:NS_final}\\
     \nabla^2 \rho' = Pe \big((\bu'-\bU)\cdot\bnabla\rho' - \bu' \cdot \bgh\big).
    \label{eqn:Advec-Diff_final}
\end{gather}
Also for an inertialess particle with no net weight, equations \eqref{eqn:linmom} and \eqref{eqn:angmom} simplify to state that the disturbance hydrodynamic force and torque on the particle are zero,
\begin{align}
\bF' = \int_{\partial\fB}\bn\cdot\bsigma' dS = \bzero,\label{eqn:force-free}\\
\bL' = \int_{\partial\fB}\br\times(\bn\cdot\bsigma') dS = \bzero.\label{eqn:torque-free}
\end{align}
In the next sections we develop the analytical approach to solving the active particle dynamics to leading order in small $Ri_v$ and $Pe$ with the aid of the reciprocal theorem.

\section{\label{app:reciprocal}Reciprocal Theorem} 
We find the particle velocities $\bU$ and $\bOmega$ by applying the generalized reciprocal theorem to the unknown flow generated by the active particle and a known auxiliary flow. The latter auxiliary flow concerns the flow induced by the rigid body motion of an inertialess passive spherical particle in a homogeneous Newtonian fluid of viscosity $\rho_{\infty} \nu$. Denoting the variables associated with auxiliary flow with an overhead bar, this flow satisfies the incompressible Stokes equations
\begin{equation}
    \nabla \cdot \bub = 0, \nabla \cdot {\bsigmab} = \bzero
    \label{eqn:auxil}
\end{equation}
and it equals the particle’s translational $\bUb$ and angular velocity $\bar{\bOmega}$ on the particle surface but vanishes far from the particle 
\begin{gather}
    \bub = \bUb + \bar{\bOmega} \times \br \,\, \text{for} \,\, \bx\in \partial\fB, 
    \label{eqn:BC_passive}\\
    \bub = \bzero \,\, \text{as} \,\, r \to \infty.
\end{gather}
The linearity of these equations and boundary conditions enforces the flow and stress fields to be linear in the particle velocities, which in terms of the six-dimensional vector $\tUb$ is $\bub = \tGb_{\tU}\cdot\tUb$, $\bsigmab = \tTb_{\tU}\cdot\tUb$. Likewise, the force and torque acting on the particle are linear in its velocity $\mathsf{\t{\bar{F}}} = \bar{\tR}_{\tF\tU} \cdot \tUb $. The expressions for the linear operators $\tGb_{\tU}$, $\tTb_{\tU}$ and the resistance tensor $\bar{\tR}_{\tF\tU}$ can be found in \cite{Happel1981}.

Following the usual procedure \cite{Happel1981, Elfring2017, Masoud2019}, we derive the reciprocal theorem by combining the equations governing the flow of an active particle \eqref{eqn:cont_final}, \eqref{eqn:NS_final} and the auxiliary flow \eqref{eqn:auxil}
\begin{equation}
    \int_{\partial \fV} \bn \cdot \bsigmab \cdot \bu' \, dS - \int_{\partial \fV} \bn \cdot \bsigma' \cdot \bub \, dS = \int_{\fV} \bf \cdot \bub \, dV.
\end{equation}
Here $\bf = Ri_v \rho' \bgh$, the normal $\bn$ points into the fluid, $\fV$ is the entire fluid volume bounded by the particle surface $\lb( \partial \fB \rb)$ on the inside and a surface in the far-field on the outside, and $\partial \fV$ denotes these bounding surfaces. The surface integrals over the latter surface vanish due to the rapid decay of the stress and velocity fields compared to the slow growth of surface area, hence we replace $\partial \fV$ with $\partial \fB$ in surface integrals. We further simplify the theorem by enforcing the boundary conditions on the surfaces of active \eqref{eqn:BC_activeparticle} and passive \eqref{eqn:BC_passive} particles as well as the force-free and torque-free constraints of active particle \eqref{eqn:force-free}, \eqref{eqn:torque-free} to derive %\textcolor{red}{We use a generalized reciprocal theorem to derive the dynamics of a force and torque free active particle moving through forced Stokes flow $\bnabla\cdot\bsigma' = \bf$, with $\bf = Ri_v \rho'\bgh$,}
\begin{align}
\tU=\bar{\tR}_{\tF\tU}^{-1}\cdot\lb[\tF_s +\tF_f \rb].
\label{eqn:vel_recip}
\end{align}
The formulas for forces are given as projections onto the known rigid-body motion auxiliary flow
\begin{align}
      \tF_s &= \int_{\partial\fB}\bu^s\cdot(\bn\cdot\tTb_{\tU}) d S,\\
      \tF_f &= \int_{\fV} \bf\cdot\tGb_{\tU} dV.
\end{align}

For a spherical particle, separating translational and rotational velocities and explicitly writing the forcing term $\bf = Ri_v\rho'\bgh$ we obtain 
\begin{align}
\bU = &-\frac{1}{4\pi} \int_{\partial\fB}\bu^s \, dS \nonumber\\
&+ \frac{1}{8\pi} \int_{\fV} \lb[\frac{\bI}{r}+\frac{\br\br}{r^3}+\frac{\bI}{3r^3}-\frac{\br\br}{r^5}\rb]\cdot(Ri_v\rho'\bgh) \, dV, \\
    \bOmega = &-\frac{3}{8\pi} \int_{\partial\fB}{\bn \times \bu^s} \, dS -\frac{1}{8\pi} \int_{\fV}\frac{\br}{r^3} \times(Ri_v\rho'\bgh)\, dV.
\end{align}
When $Ri_v=0$ the particle moves through a homogeneous Newtonian fluid where the dynamics of a squirmer are well known \cite{Stone1996}, $\bU = \bU_N = \bp$ and $\bOmega = \bzero$, so the change in velocities due to an inhomogeneous fluid simplify to
\begin{align}
\bU-\bU_N &= \frac{Ri_v}{8\pi}\bgh\cdot\int_{\partial\fB} \bigg[(\bI+\bn\bn) \int_1^\infty \rho' rdr\nonumber\\
&\quad+(\bI/3 - \bn\bn) \int_1^\infty \frac{\rho'}{r} dr\bigg] dS,\label{eqn:DeltaU_recip}\\
\bOmega &= \frac{Ri_v}{8\pi}\bgh\times\int_{\partial\fB} \bn \lb(\int_1^\infty \rho'dr\rb) dS \nonumber \\ &= -\frac{Ri_v}{8\pi}\int_{\partial\fB} \bn \times\lb(\int_1^\infty \rho' \bgh dr\rb) dS.
\label{eqn:Omega_recip}
\end{align}
where we've used the fact that $\br/r = \bn$ on a sphere. To make analytical progress in evaluating these integrals, we find the density disturbance assuming relativey weak advection.

\section{\label{app:analysis}Analysis}
When advection is weak, density transport \eqref{eqn:Advec-Diff_final} is dominated by diffusion near the sphere; however, no matter how small $Pe$ there is always a distance, characterized by the screening length $l_{\rho} \sim 1/Pe$, beyond which advection terms become on the order of diffusion and this indicates a regular expansion in $Pe$ is not uniformly valid \cite{Leal2007}. To solve for $\rho'$ one must carefully consider both regions where $1 < r < l_{\rho}$ and where $r > l_{\rho}$.

\begin{figure}[t!]
    \centering
    \includegraphics[scale = 0.5]{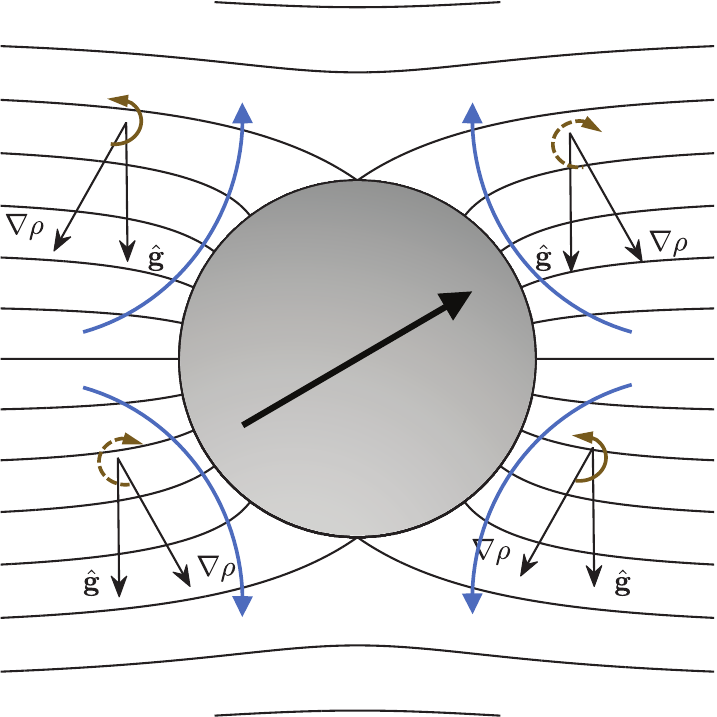}
    \caption{Baroclinically induced flow and vorticity around a spherical passive or active particle. Black lines denote the isopycnals while brown and blue curved arrows denote, respectively the vorticity and flow directions.}
    \label{fig:fig3}
\end{figure}

In the limit that $Pe\rightarrow 0$, the density satisfies a Laplace equation, $\nabla^2\rho'=0$, which when coupled with no-flux conditions on the boundary yields a solution that decays at infinity (and is thus uniformly valid)
\begin{align}
\rho' = \frac{\br}{2r^3}\cdot\bgh.
\label{eqn:density0}
\end{align}
With this density field the integrals in \eqref{eqn:DeltaU_recip} and \eqref{eqn:Omega_recip}, are zero by symmetry, meaning $\bU-\bU_N = \bzero$ and $\bOmega = \bzero$ in the limit $Pe\rightarrow 0$. In particular, the symmetry of the disturbance density field implies that the induced baroclinic vorticity $\left( \propto \nabla \rho' \times {\hat\bg} \right)$ never exerts any torque on the particle nor rotates the particle (see Fig.~\ref{fig:fig3} for a schematic). Baroclinic flow is induced by a misalignment of surfaces of constant density (isopycnals) from the horizontal, arising here due to the no-flux boundary condition on the particle surface. The computation of this flow is however bypassed in the application of the reciprocal theorem to compute the changes in translational and angular velocity.

Corrections to the leading order disturbance density in \eqref{eqn:density0} requires resolution of the velocity field. To do so we will assume that the buoyancy force is relatively weak, $Ri_v\ll 1$. Here we note that an expansion of the forced Stokes equations for $Ri_v\ll 1$ is also singular, as no matter how small $Ri_v$, buoyancy forces become comparable to viscous forces at a distance from the particle characterized by the stratification screening length $l_s$ (formally defined below). Asymptotic solutions at higher order depend on the relative order of magnitude of the two asympototic parameters, $Pe$ and $Ri_v$. Framed another way, there are three regions in the asymptotic solution, and so density transport and hence buoyancy forces far from the particle depend on $l_s/l_{\rho}$. 

As we are interested in the buoyancy induced velocity changes, for convenience we will refer to the region where buoyancy forces are greater than viscous forces, $r > l_s$, as the `outer' region, and conversely when $1 < r <  l_s$ the `inner' region, but carefully account for the effect of advection on density. Specifically, we note than when $l_s < l_{\rho}$, the outer region is comprised of two domains, the first ($l_{\rho} > r > l_s $) where density disturbances (and thus buoyancy forces) are governed by diffusion, beyond which ($r > l_{\rho}$) they are governed by advection. Alternatively, when $l_s > l_{\rho}$, buoyancy forces are governed by advection in the outer region. In the next sections, we find the disturbance density and with that the dynamics of the active particle in the two asymptotic cases where the domains are well separated, $l_s/l_{\rho} \gg 1$ and $\ll 1$.

\subsection{$l_s \ll l_{\rho}$ ($Pe \ll Ri_v^{1/3}$)}
%We find the correction to disturbance density in \eqref{eqn:density0} and evaluate the integral in \eqref{eqn:Omega_recip} separately in inner and outer zone.

We first consider the case where $l_s \ll l_{\rho}$. Assuming a uniformly valid regular perturbation expansion, a correction to the density perturbation given by \eqref{eqn:density0}, that scales as $\sim f(\br)Pe$, would give an angular velocity $\bOmega \sim O\left( Ri_v Pe \right)$ using \eqref{eqn:Omega_recip}, provided $\int_1^\infty fdr=O(1)$; however, the integral of the density perturbation diverges indicating such a correction is valid only in the inner region $\left( r < l_s \right)$ and whose dominant contribution to $\bOmega$ comes from the overlap regime where $r\sim O(l_s)$. To establish this more precisely, we know the disturbance flow far from a force-free particle $\left( 1 \ll r < l_s \right)$ is that of a force dipole so that $\bu' \sim 1/r^2$ (for small $Ri_v$). Using this, we can estimate the change in density at finite $Pe$ by balancing diffusion with advection of background density
\begin{align}
\nabla^2 \rho' \sim Pe \bu' \cdot \bgh,
\label{eqn:densityPe}
\end{align}
where the advection of the perturbation term is neglected as $\bnabla\rho'$ decays more rapidly than $\bu'$. From \eqref{eqn:densityPe} we find the change in density due to advection $\sim f(\bn)\,Pe$ (where $f$ is $O(1)$ and independent of $r$). This means that the density change due to advection exceeds the leading order term \eqref{eqn:density0} at distances $r > Pe^{-1/2}$. Substitution of $\rho'$ into \eqref{eqn:Omega_recip} yields a leading order term $\bOmega \sim Ri_v Pe \, r, \ r\rightarrow \infty$, indicating this inner solution is not uniformly valid. This divergence is cut off at the screening length $r \sim l_s$ due to rapid decay of density beyond this distance and it implies the dominant contribution to $\bOmega$ comes from the overlap region $ r \sim l_s $. Using the advection density scaling, we obtain the stratification screening length by balancing the buoyancy forces with the viscous forces at $r \sim l_s$
\begin{align}
\nabla^2 \bu' \sim Ri_v \rho'\bgh.
\end{align}
With $\rho'\sim Pe$ we find $l_s \sim \left( Ri_v Pe \right)^{-1/4}$. With this scaling we find that $\bOmega \sim \left( Ri_v Pe \right)^{3/4}$ and, because this contribution is due entirely to the overlap region common to both inner and outer regions, we conclude that the orientational dynamics, $\bOmega$, is be determined entirely by the outer region alone to leading order.

%This far-field contribution can be found by evaluating $\bOmega$ in the outer region alone $\left( r > l_s \right)$ as $r \sim l_s \gg 1$ is common to both inner and outer regions. Said differently, the dominant contribution to $\bOmega$ comes from the outer region, and this is often revealed through the divergence of $\bOmega$ with $r$ when evaluated using the inner region quantities. We next confirm the scaling of $\bOmega$ with $Ri_v$ and $Pe$ by evaluating it in the outer region.

Next, we derive equations valid in the far-field $\left( r \gg 1 \right)$. Here, we add appropriate singular forcing terms to \eqref{eqn:cont_final}-\eqref{eqn:Advec-Diff_final} to account for the presence of the force-free active particle to leading order in the far field. Specifically, a force dipole of strength $\bS$ and a degenerate quadrupole of strength $\bQ$, for the disturbance flow, while the density disturbance generated by the particle has character of concentration-dipole of strength $\bD$. Adding the singular forcing to \eqref{eqn:NS_final} and \eqref{eqn:Advec-Diff_final}, neglecting products of disturbance variables, and approximating $\bU$ as $\bU_N$ we have
\begin{gather}
\nabla \cdot \bu' = 0,
\label{eqn:Cont-outer}\\
-\bnabla p' + \nabla^2 \bu' = Ri_v \rho' \bgh - \bS \cdot \bnabla \delta \left( \br\right) - \bQ \nabla^2 \delta \left( \br\right),\\
\nabla^2 \rho'+Pe \bU_N \cdot \bnabla \rho' + Pe \bu' \cdot \bgh =  -\bD \cdot \bnabla \delta \left( \br \right).
\label{eqn:densityTrans-outer}
\end{gather}
We note that the only assumptions made in deriving these equations (other than the form of the forcing) have been linearizing by discarding the nonlinear advection of the disturbance density by the disturbance flow, and approximating the particle velocity with its value in homogeneous fluid. The former assumption holds quite generally in the far-field due to the smallness of the disturbance variables, while the latter one holds for weak density gradients $Ri_v \ll 1$. More generally, these far-field equations are valid at any value of $l_s/l_{\rho}$ and also at both small and large $Pe$ as no assumption on these quantities was made while deriving them. Also, the coefficients of the singularites, $\bS$, $\bQ$, and $\bD$, are generally functions of $Ri_v$ and $Pe$, but to leading order, they can be treated as independent of these dimensionless numbers, $\bS = 2 \pi \beta \left( 3 \bp\bp - \bI \right) $, $\bQ = - 2 \pi \bp$ and $\bD = -2 \pi \bgh$.

%We rescale the variables in outer zone $\left( l_{\rho} >  r > l_s \gg 1 \right)$ to demonstrate that the buoyancy forces balance the viscous forces and that diffusion primarily governs the density transport. 

For convenience, we rescale such that rescaled variables, denoted with a tilde, are $O(1)$ in the matching region $\left( r \sim l_s\right)$. Consequently, $\tilde{r} = r \left( Ri_v Pe \right)^{1/4}  $, $\rhot'=\rho'/ Pe$, $\but'=\bu'/\left( Ri_v Pe\right)^{1/2} $ and $\pt'=p'/\left( Ri_v Pe \right)^{3/4}$. In terms of the rescaled variables, the governing equations in the far-field \eqref{eqn:Cont-outer}-\eqref{eqn:densityTrans-outer} simplify to
\begin{gather}
    \tilde{\bnabla} \cdot {\tilde\bu}' = 0,\label{eqn:cont-rescaled}\\
    - \tilde{\bnabla} \tilde{p}' + \tilde{\nabla}^2 {\tilde\bu}' \qquad\qquad\qquad\qquad\qquad\qquad\qquad\qquad\nonumber\\
    \qquad = \tilde{\rho}' \bgh - \bS \cdot \tilde{\bnabla} \delta \left( \tilde\br \right) - \left( Ri_v Pe\right)^{1/4} \bQ\tilde{\nabla}^2 \delta \left( \tilde\br \right),\\
 \tilde{\nabla}^2 \tilde{\rho}'   + \frac{Pe^{3/4}}{Ri_v^{1/4}} \, \bU_N \cdot \tilde{\bnabla} \tilde{\rho}' + {\tilde\bu'} \cdot \bgh =  \sqrt{\frac{Ri_v}{Pe}} \, \bD \cdot \tilde{\bnabla} \delta \left( \tilde\br \right).
    \label{eqn:density-rescaled}
\end{gather}

Similarly, the angular velocity in \eqref{eqn:Omega_recip} in terms of rescaled variables becomes
\begin{equation}    
% \bOmega = -\left( Ri_v Pe \right)^{3/4}\frac{1}{8\pi} \int_{\tilde{\fV}}\frac{\brt}{\rt^3} \times(\rhot'\bgh)\, d\tilde{V}.\label{eqn:Omega-rescaled}
 \bOmega = \frac{(Ri_v Pe)^{3/4}}{8\pi}\bgh\times\int_{\partial\fB} \bn \lb(\int_{1/l_s}^\infty \rhot'd\rt\rb) dS.
\label{eqn:Omega-rescaled}
\end{equation}
The rescaling shows the explicit dependence of $\bOmega$ on $Ri_v$ and $Pe$, found for the inner region truncating at $r \sim l_s$. The outer solution $\tilde{\rho}'$ here although is strictly valid in the far-field of the particle $\left( r \gg 1 \right)$, it can be integrated over the whole fluid domain $\left( 1 < r < \infty \,\, \text{or} \,\, 1/l_s < \tilde{r} < \infty  \right)$. Such integration of the composite solution ($=$ inner $-$ matching $+$ outer solution) instead makes sense as this solution holds throughout the entire fluid domain. But since the outer solution component of the composite solution solely determines leading order rotation, it can instead be integrated over the whole domain.

%The remaining integral brings in the dependence on the squirming ratio $\beta$ and the particle orientation relative to gravity $\psi = \arccos\left( \bp \cdot \hat\bg \right)$.

We solve Eqs.~\eqref{eqn:cont-rescaled}-\eqref{eqn:density-rescaled} in Fourier space to find the Fourier transform of density. We denote the Fourier transform of a function $\tilde{f} \left( \tilde\br\right)$ by $\hat{f} \left( \bk\right)$, with the Fourier and inverse Fourier transforms following the definition
\begin{equation}
    \hat{f} \left( \bk \right) = \int{\tilde{f} \left( \tilde\br\right) e^{-i \bk \cdot \tilde\br }} d\tilde\br, \thickspace \tilde{f} \left( \tilde\br\right) = \frac{1}{8\pi^3} \int{ \hat{f} \left( \bk \right)  e^{i \bk \cdot \tilde\br } d\bk}.
\end{equation}

While we can solve the system above for the density in Fourier space for arbitary values of $Pe$ and $Ri_v$, to facilitate an inverse transform we expand $\hat{\rho}' \left( \bk \right)$ assuming $(Ri_v Pe )^{1/4}\ll 1$ $(l_s\gg 1)$, and $Pe^{3/4}/Ri_v^{1/4}\ll 1$ $(l_s/l_\rho \ll 1)$, such that 
\begin{multline}
\hat{\rho}' \left( \bk \right) = \hat{\rho}'_{00} \left( \bk \right) +  \left(Ri_v Pe \right)^{1/4} \hat{\rho}'_{01} \left( \bk \right) \\+ Pe^{3/4}/Ri_v^{1/4} \hat{\rho}'_{10} \left( \bk \right) + ...,
\end{multline}
where
\begin{align}
    \hat{\rho}_{00}' = \,\,&\frac{6i\beta \pi \left( \bp \cdot \bk \right) \left( \left( \bk \cdot \bgh \right) \left( \bp \cdot \bk \right) - \left( \bp \cdot \bgh \right) k^2 \right)}{k^6 + k^2 - \left( \bk \cdot \bgh \right)^2} \nonumber \\ &- \sqrt{\frac{Ri_v}{Pe}} \frac{2 \pi i k^4 \left( \bk \cdot \bgh \right)}{k^6 + k^2 - \left( \bk \cdot \bgh \right)^2}, \label{eqn:dens00L}\\
    \hat{\rho}_{01}' = \,\,&\frac{2 \pi k^2 \left( - k^2 \left( \bp \cdot \bgh \right) + \left( \bk \cdot \bgh \right) \left( \bp \cdot \bk \right) \right) }{k^6 + k^2 - \left( \bk \cdot \bgh \right)^2},\\
    \hat{\rho}_{10}' = \,\,&- \frac{ 6 \pi \beta k^4 \left( \bp \cdot \bk \right)^2 \left( \left( \bk \cdot \bgh \right) \left( \bp \cdot \bk \right) - k^2 \left( \bp \cdot \bgh \right) \right)  }{ \left( k^6 + k^2 - \left( \bk \cdot \bgh \right)^2 \right)^2 } \nonumber \\ &+ \sqrt{\frac{Ri_v}{Pe}} \frac{ 2 \pi k^8 \left( \bk \cdot \bgh \right) \left( \bp \cdot \bk \right) }{\left( k^6 + k^2 - \left( \bk \cdot \bgh \right)^2 \right)^2}.
\end{align}

To facilitate the evaluation of the angular velocity, it is useful to extend the volume of integration in \eqref{eqn:Omega-rescaled} to include the particle volume, $\int_{1/l_s}^{\infty}{\tilde{\rho}' d\tilde{r}} \to \int_{0}^{\infty}{\tilde{\rho}' d\tilde{r}}$. In this manner we can transform the integral from real space to Fourier space using the convolution theorem and evaluate it with the leading order density $\hat{\rho}'_{00}$ to derive
\begin{align}
    {\bOmega} &=  -\frac{\left( Ri_v Pe \right)^{3/4}}{16 \pi^3}\bgh\times  \int \hat{\rho}' \left( \bk \right)\frac{i\bk}{k^2}   d\bk.
    \nonumber\\
    &= \frac{\beta \left( Ri_v Pe \right)^{3/4}}{8} \left( 2 E_E\left(\frac{1}{\sqrt{2}}\right) - E_K \left(\frac{1}{\sqrt{2}}\right)\right) \nonumber \\ & \qquad\qquad\qquad\qquad\qquad\qquad\,\,\,\,\,\,\,\, \left( \bp\cdot \hat\bg \right) \left( \bp\times \hat\bg \right), \label{eqn:Omega-result1}
\end{align}
where $E_K$ and $E_E$, respectively, are the complete elliptic integrals of the first and second kind. Including particle volume in integral volume does not make the integral divergent even though the density $\tilde{\rho}' \left( \sim 1/ \tilde{r}^2 \right)$ diverges with $\tilde{r}$ as $\tilde{r} \to 0$ due to the concentration-dipole forcing. This is because the density $\tilde{\rho}' \sim \sqrt{\frac{Ri_v}{Pe}} \frac{\brt \cdot \bgh}{2 \rt^3}$ is top-down mirror-symmetric and hence vanishes upon overall integration. This is also apparent in Fourier space where the density due to concentration-dipole (the term proportional to $\sqrt{\frac{Ri_v}{Pe}}$ in \eqref{eqn:dens00L}) does not induce rotation. But such inclusion of particle volume induces an (insignificant) error of  $\left( Ri_v Pe \right)^{3/4} \int_{\tilde{\fV}_p} \frac{\brt}{\rt^3} \times\bgh \, \rhot'\, d\tilde{V} \sim \left( Ri_v Pe \right)^{3/4} \int_{\tilde{\fV}_p} \frac{\brt}{\rt^3} \times\bgh \left( \rhot' -  \sqrt{\frac{Ri_v}{Pe}} \frac{\brt \cdot \bgh}{2 \rt^3} \right) d\tilde{V} \sim \left( Ri_v Pe \right)^{3/4} \int_{\tilde{\fV}_p} \frac{1}{\rt^2} \, d\tilde{V} \sim \left( Ri_v Pe \right)^{3/4} \rt \sim Ri_v Pe $ as $ \rhot' -  \sqrt{\frac{Ri_v}{Pe}} \frac{\brt \cdot \bgh}{2 \rt^3} \sim O\left(1 \right)$ and $\rt \sim O\left( Ri_v Pe \right)^{1/4}$ near the particle.

\begin{figure}[t!]
    \centering
    \includegraphics[scale = 0.55]{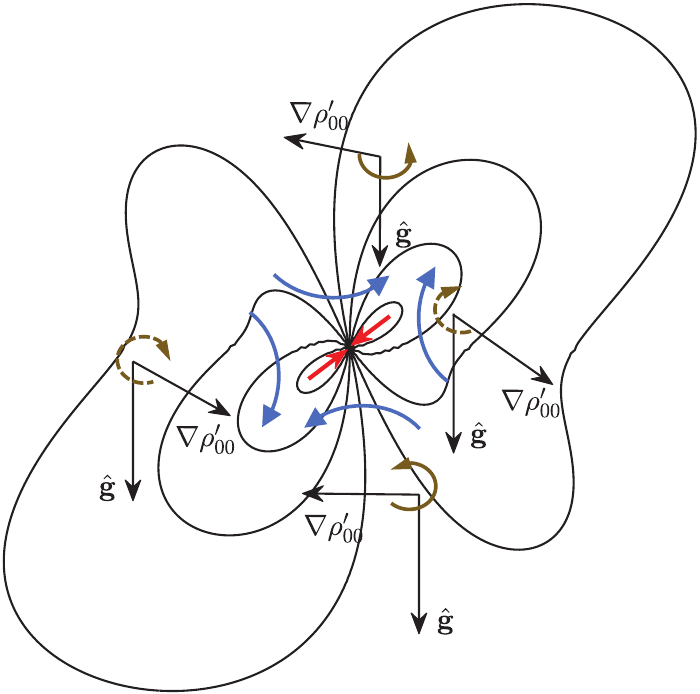}
    \caption{Baroclinically induced flow and vorticity in the far-field of a puller aligned with respect to gravity at $\psi = 3\pi/4$ and in the limit $Ri_v \ll Pe \ll Ri_v^{1/3}$. Black lines denote the isopycnals while brown and blue curved arrows denote, respectively the vorticity and flow directions.}
    \label{fig:fig4}
\end{figure}

A finite rotation at this order can be understood by analyzing the far-field isopycnals and baroclinically induced flow. See Fig.~\ref{fig:fig4}. Isopycnals here are found by performing the inverse Fourier transform of $\hat{\rho}'_{00}$ using the {\texttt{ifftn}} command in MATLAB. They lack any top-down or left-right symmetry, unlike the isopycnals in near-field, and hence the ensuing the baroclinic flow should rotate the particle. But the direction of rotation is not so obvious, as like in the near-field, the far-field baroclinic flow in different quadrants rotates the particle in different directions, clockwise or counter-clockwise. Flows inducing opposite sense of rotation are of equal magnitude in near-field but of different magnitude in the far-field and hence result in finite particle rotation. %\textcolor{red}{(not exactly a great physical argument, you basically say, well it isn't symmetric, so there should be some rotation, but have no idea the direction...)}\\

We next find the velocity changes $\Delta \bU = \bU - \bU_N$ by evaluating the integral in \eqref{eqn:DeltaU_recip}. As the kernel against density in this integral decays slower than that appearing in \eqref{eqn:Omega_recip}, we expect that this integral diverges with $r$ like the latter one when evaluated with the advective corrections to diffusive density $\rho' \sim f\left( \bn \right)Pe$. This divergence again implies the dominant contribution to $\Delta \bU$ comes from the overlap regime $r \sim O\left( l_s \right)$ and can be found by evaluating it solely in the outer region.

In the outer zone, we simplify the integral in velocity changes \eqref{eqn:DeltaU_recip} before evaluating it. We retain only the slowest decaying terms in the kernel multiplying the density, and rescale the variables in the usual manner to derive
\begin{equation}
    \Delta \bU = \frac{\left( Ri_v Pe \right)^{1/2}}{8 \pi} \bgh \cdot \int_{\partial\fB} \lb( \bn \bn + \bI \rb) \lb(\int_{1/l_s}^\infty \rhot' \rt d\rt\rb) dS.
    \label{eqn:DeltaU-rescaled}
\end{equation}
We then extend the volume of integration to include the particle volume and transform the integral from real space to Fourier space to derive
\begin{equation}
\Delta \bU = \frac{ \lb( Ri_v Pe \rb)^{1/2} }{8 \pi^3} \bgh \cdot \int \rhoh' \lb( \bk \rb) \frac{1}{k^2} \lb( \bI - \frac{\bk \bk}{k^2} \rb) d \bk.
\label{eqn:DeltaU-Fourier}
\end{equation}
It is now apparent that the velocity changes from the outer region occur at $O\lb( Ri_v Pe \rb)^{1/2}$ but a detailed evaluation of the integral reveals that the velocity changes vanish at this order but are finite at next order $O\lb( Ri_v^{1/4} Pe^{5/4}, \lb( Ri_v Pe \rb)^{3/4} \rb)$. Also including particle volume in the integral volume does not make the integral divergent as the density near the particle is still top-down mirror-symmetric and hence vanishes upon overall integration, but it induces an insignificant error of $O \lb( Ri_v Pe \rb)$ that restricts the validity of $\Delta \bU$ to $Ri_v^3 \ll Pe \ll Ri_v^{1/3}$.

We find the velocity changes by evaluating the integral in \eqref{eqn:DeltaU-Fourier}. At leading order, the density $\hat{\rho}'_{00}$ (or its real space counterpart) has no symmetry for arbitrary particle orientations but it still induces zero velocity changes $\Delta \bU = \bzero$. In the relevant vertical or horizontal orientations, however, the density in real space is top-down mirror-symmetric or odd in $\br \cdot \bgh$ as the density in Fourier space is odd in $\bk \cdot \bgh$. See \eqref{eqn:dens00L}. This symmetry is caused by the concentration-dipole forcing $\bD \cdot \tilde{\nabla} \delta \lb( \brt\rb)$ or the advection of background density by the Stresslet flow $\but' \cdot \bgh$. It means that, relative to the background fluid, the displaced fluid at the top of the particle is heavier and that at the bottom of the particle is lighter by same amount or vice-versa. Any weight of the overall displaced fluid acts as buoyancy and alters the particle velocity but the zero weight of the displaced fluid at this order does not change the particle velocity. The aforementioned mirror-symmetry is however lost at next order due to advection of disturbed density with the particle velocity $\bU_N \cdot \tilde{\nabla} \tilde{\rho}'$ or the advection of background density by the degenerate quadrupole flow $\but' \cdot \bgh$. Hence, finite velocity changes are expected at this order and can be found by evaluating \eqref{eqn:DeltaU-Fourier} with the first order densities $\hat{\rho}'_{01}$ and $\hat{\rho}'_{10}$
\begin{widetext}
\begin{multline}
\Delta \bU = - \frac{\beta Ri_v^{1/4} Pe^{5/4} }{2464} E_K\left( \frac{1}{\sqrt{2}}\right) \left( 31 \left( \bp \cdot \hat\bg \right)^2 \bp - 15 \bp - 21 \left( \bp \cdot \hat\bg \right)^3 \hat\bg - 35 \left( \bp \cdot \hat\bg \right) \hat\bg \right)\\
- \frac{\left( Ri_v Pe \right)^{3/4}}{120}\left( 2 E_E \left(\frac{1}{\sqrt{2}}\right) - E_K\left(\frac{1}{\sqrt{2}}\right)  \right) \left( 7 \bp + 15 \left( \bp \cdot \hat\bg \right) \hat\bg \right).
\end{multline}
\end{widetext}
Then vertical swimming $\left( \bp = \pm \hat\bg \right)$ occurs at velocity
\begin{multline}
    \bU = \left\{ 1 + \frac{5 \beta Ri_v^{1/4} Pe^{5/4} }{308} E_K\left( \frac{1}{\sqrt{2}} \right) \right. \\ \qquad \left.- \frac{11 \left( Ri_v Pe \right)^{3/4}}{60} \left( 2 E_E \left( \frac{1}{\sqrt{2}}\right) - E_K \left( \frac{1}{\sqrt{2}} \right) \right) \right\} \bp,
\end{multline}
which is in line with that reported earlier \cite{Shaik2021} in the restrictive limit $Ri_v \ll Pe \ll Ri_v^{1/3}$ where the term at $O\left( Ri_v Pe \right)^{3/4}$ is negligible. Speed changes in this configuration are indifferent to whether the swimming is up or down the density gradients. Also, pullers swimming along this (steady state) direction speed up relative to their speed in the absence of density gradients if
\begin{equation}
    \frac{300}{3388} \frac{E_K\left( \frac{1}{\sqrt{2}} \right)}{ 2 E_E\left( \frac{1}{\sqrt{2}} \right) - E_K \left( \frac{1}{\sqrt{2}} \right) } \beta \sqrt{\frac{Pe}{Ri_v}} > 1,
\end{equation}
else slow down. This expands on the previous observation of pullers only speeding up \cite{Doostmohammadi2012, More2020, Shaik2021} and suggests a possibility of better control over particle speed in density gradients by controlling the appropriate dimensionless numbers. On the other hand, horizontal swimming $\left( \bp \perp \hat\bg \right)$ occurs at velocity
\begin{multline}
    \bU = \left\{ 1 + \frac{15 \beta Ri_v^{1/4} Pe^{5/4}}{2464} E_K \left( \frac{1}{\sqrt{2}} \right) \right. \\ \left. - \frac{7 \left( Ri_v Pe \right)^{3/4}}{120} \left( 2 E_E \left( \frac{1}{\sqrt{2}} \right) - E_K \left( \frac{1}{\sqrt{2}} \right) \right) \right\} \bp,
\end{multline}
hence pushers swimming along this (steady state) direction always slow down.

\begin{figure}[t]
    \centering
    \includegraphics[scale = 0.4]{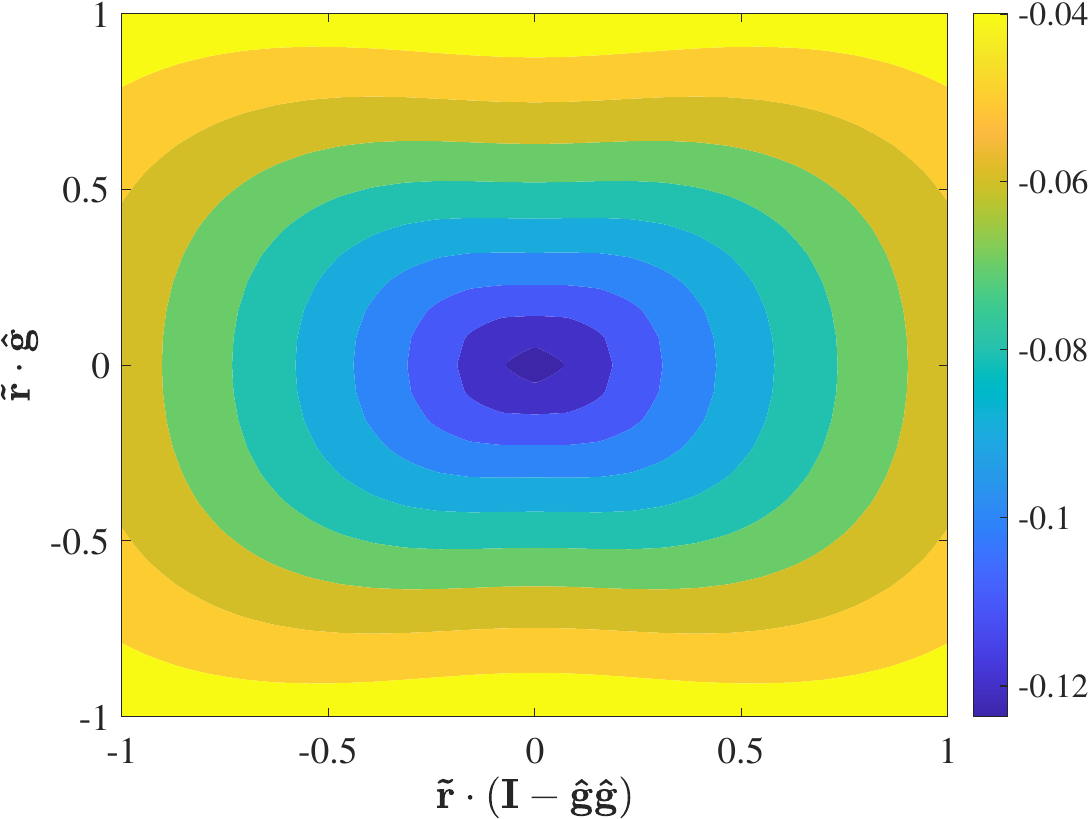}
    \caption{First order density disturbance $\tilde \rho'_{10}$ in the far-field of a vertically upwards swimming puller for $Ri_v \ll Pe \ll Ri_v^{1/3}$. Figure adapted and modified from Ref.~\cite{Shaik2021} with permission from the American Institute of Physics.}
    \label{fig:fig5}
\end{figure}

Speed changes along steady-state swimming direction can be understood by analyzing the density of the displaced fluid. Neutrally buoyant active particles displace a fluid of volume $O\lb( b/(Ri_v Pe)^{1/4} \rb)$, where $b \propto \beta$ is the stresslet strength \cite{Shaik2020}. This corresponds to a fluid around the particle of spatial extent $r \sim O\lb( Ri_v Pe \rb) ^ {-1/12}$  or $\rt \sim O\lb( Ri_v Pe \rb)^{1/6} \ll 1$. Disturbed density of the fluid displaced by a vertically upwards swimming puller is negative for $Ri_v \ll Pe \ll Ri_v^{1/3}$, see Fig.~\ref{fig:fig5}, which means that a lighter fluid is displaced to a region otherwise occupied by a heavier fluid. This lighter fluid experiences an upward buoyancy that in turn speeds up the puller.

\subsection{$l_s \gg l_{\rho}$ ($Pe \gg Ri_v^{1/3}$)}

In the other regime, $l_s \gg l_{\rho}$. Far from the particle $1\ll l_\rho < r < l_s$, now advection dominates diffusion, and so we estimate the density perturbation by balancing the advection of disturbed density with the advection of background density
\begin{align}
Pe \bU \cdot \nabla \rho' \sim Pe \bu' \cdot \bgh \
\end{align}
Again using $\bu' \sim 1/r^2$ we obtain $\rho' \sim 1/r$. Substitution of $\rho'$ into \eqref{eqn:Omega_recip} leads to $\bOmega \sim Ri_v \ln r \ r\rightarrow \infty$ which again diverges (albeit a weaker logarithmic divergence) and this divergence is cut off at distances $r\sim l_s$. In order to determine the stratification screening length we again balance viscous forces and buoyancy forces, now with $\rho' \sim 1/r$, to obtain $l_s \sim Ri_v^{-1/3}$. This implies a contribution to the angular velocity from the overlap regime that scales as $\bOmega \sim O\left( Ri_v \ln{Ri_v} \right)$. What we show below, by directly calculating the far-field solution is that the outer region leads to a correction $\bOmega \sim O\left(Ri_v\right)$.

%Density could also decay slowly beyond the stratification screening length in vertical jets or horizontal wakes, like that reported in the disturbance of a passive particle \cite{Varanasi2022a}. 

As in the previous section we solve the governing equations in the far-field $r\gg 1$, \eqref{eqn:Cont-outer}-\eqref{eqn:densityTrans-outer}, again rescaled such that rescaled variables are $O(1)$ in the matching region $\left( r \sim l_s\right)$. In this case, $\tilde{r} = Ri_v^{1/3} r$, $\rhot'=\rho'/ Ri_v^{1/3}$, $\but'=\bu'/Ri_v^{2/3} $ and $\pt'=p'/Ri_v$ such that
\begin{gather}
    \tilde{\bnabla} \cdot {\tilde\bu}' = 0,
    \label{eqn:cont-rescaled2}\\
    - \tilde{\bnabla} \tilde{p}' + \tilde{\nabla}^2 {\tilde\bu}' + \tilde{\rho}' \bgh + \bS \cdot \tilde{\bnabla} \delta \left( \tilde\br \right) + Ri_v^{1/3} \bQ\tilde{\nabla}^2 \delta \left( \tilde\br \right) = \bzero,\\
    - \bU_N \cdot \tilde{\bnabla} \tilde{\rho}' + {\tilde\bu}' \cdot \bgh = \frac{Ri_v^{1/3}}{Pe} \, \tilde{\nabla}^2 \tilde{\rho}' + \frac{Ri_v^{2/3}}{Pe} \, \bD \cdot \tilde{\bnabla} \delta \left( \tilde\br \right).
    \label{eqn:density-rescaled2}
\end{gather}
Similarly, the angular velocity in \eqref{eqn:Omega_recip} in terms of rescaled variables becomes
\begin{equation}
%    {\bOmega} = - \frac{Ri_v}{8\pi} \int_{{\tilde{V_f}}} {\tilde{\rho}' \be_3 \cdot {\tilde\bg}_{\Omega}} \tilde{dV}.
%    \label{eqn:Omega-rescaled2}  
 \bOmega = \frac{Ri_v}{8\pi}\bgh\times\int_{\partial\fB} \bn \lb(\int_{1/l_s}^\infty \rhot'd\rt\rb) dS.
    \label{eqn:Omega-rescaled2}  
\end{equation}
Here again the rescaling itself reveals the dependence of $\bOmega$ on $Ri_v$.

\begin{figure*}[t!]
\centering
\subfloat{\includegraphics[scale = 0.44]{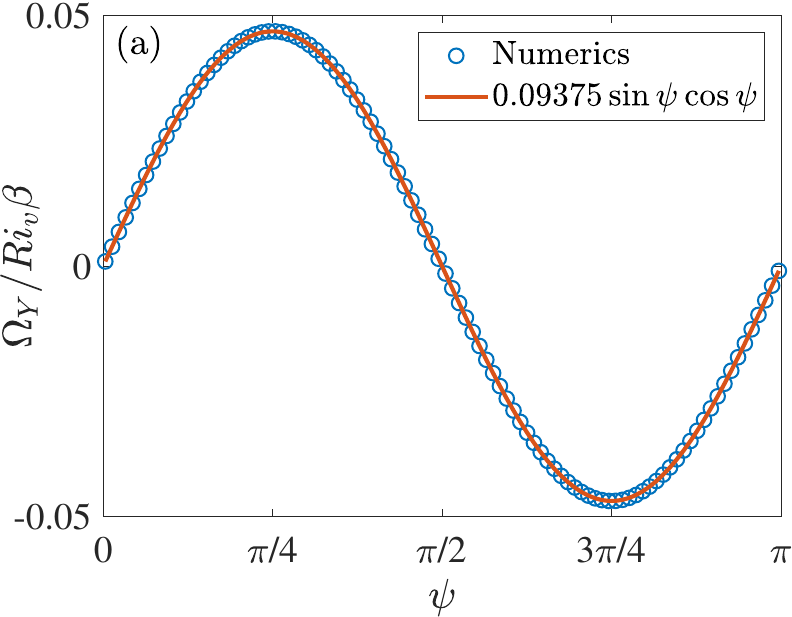}\label{fig:fig6a}}
\subfloat{\includegraphics[scale = 0.44]{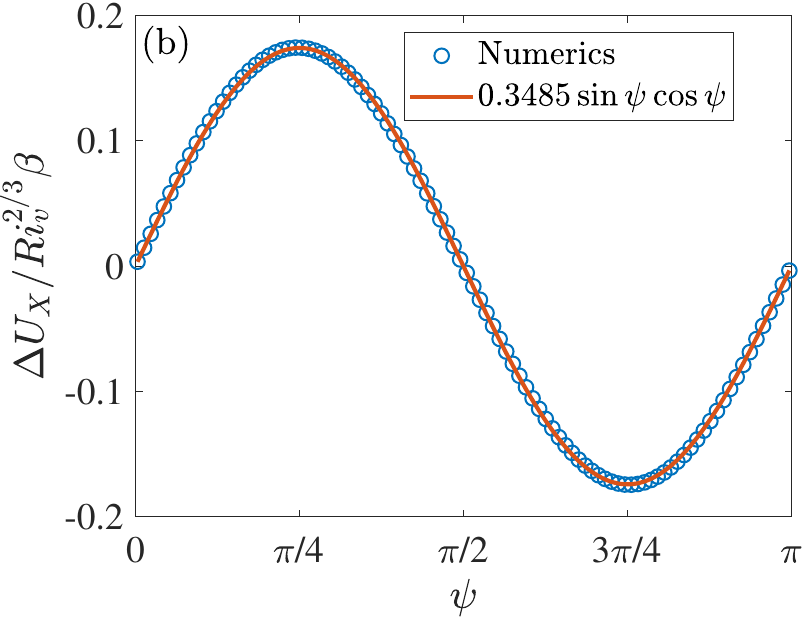}\label{fig:fig6b}}
\subfloat{\includegraphics[scale = 0.44]{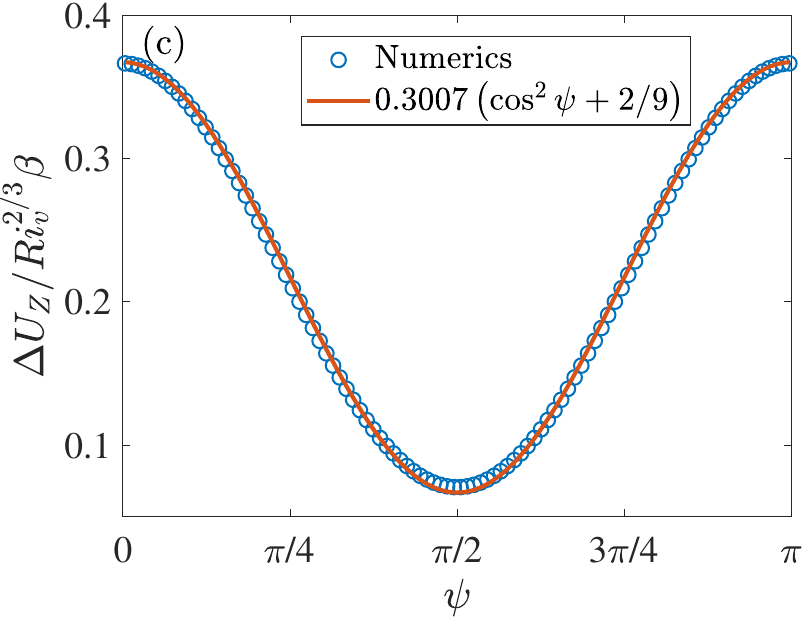}\label{fig:fig6c}}
\caption{\label{fig:fig6} Angular velocity (a) and velocity changes (b, c) in the particle aligned coordinate system for $Pe \gg Ri_v^{1/3}$. Symbols denote the numerical integration result while the lines represent the fit.}
\end{figure*}

We solve Eqs.~\eqref{eqn:cont-rescaled2}-\eqref{eqn:density-rescaled2} in Fourier space to find the Fourier transform of density $\rhoh' \left( \bk \right)$, from which we extract the leading order density $\rhoh'_{00}  \lb( \bk \rb)$ by setting two small parameters $Ri_v^{1/3}$, $Ri_v^{1/3}/Pe$ to zero 
\begin{equation}
    \hat{\rho}'_{00} = - \frac{6 \pi \beta \left( \bp \cdot \bk\right)  \left( \left( \bk \cdot \bgh \right) \left( \bp \cdot \bk \right)   - k^2 \left( \bp \cdot \bgh \right) \right)}{k^4 \left( \bp \cdot \bk\right) + ik^2 - i \left( \bk \cdot \bgh \right)^2 }.
\end{equation}
We use this leading order density to evaluate the angular velocity by transforming the integral in \eqref{eqn:Omega-rescaled2} from real space to Fourier space
\begin{equation}
    {\bOmega} = - \frac{Ri_v}{16 \pi^3} \bgh \times \int{ \rhoh' \lb( \bk \rb) \frac{i\bk}{k^2} } d\bk.
    \label{eqn:Omega-highPe-Fourier}
\end{equation}
We evaluate this integral numerically using {\texttt{integral3}} command in MATLAB and it unsurprisingly reveals that ${\bOmega}$ is non-zero only in the direction $\bp\times\bgh$. The magnitude of the rotation is linear in $\beta$ due to similar linear dependence of the leading order density on the Stresslet $\bS$. Motivated by \eqref{eqn:Omega-result1} we see here that data fits 
\begin{equation}
{\bOmega} =  c Ri_v \beta \left( \bp\cdot \hat\bg \right) \left( \bp\times\bgh \right),
\label{eqn:Omega-highPe}
\end{equation}
where the constant of proportionality, $c \approx 0.09375$ as shown in Fig.~\ref{fig:fig6a}.

A word of caution on evaluating the integral in \eqref{eqn:Omega-highPe-Fourier} in the traditional gravity-aligned coordinate system: while the overall integral converges, the individual integrals comprising $\bOmega$ diverge. This divergence is possibly due to any slow decay of density in the jets and wakes surrounding the particle, like that reported in the vicinity of a passive particle \cite{Varanasi2022a}. Both individual and overall integrals however converge in the more convenient particle-aligned coordinate system $\left\{ \be_X, \be_Y, \be_Z \right\} = \left\{ \frac{\lb( \bp \times \bgh \rb) \times \bp}{ \left| \lb( \bp \times \bgh \rb) \times \bp \right| }, \frac{\bp \times \bgh }{ \left| \bp \times \bgh \right| }, \bp \right\}$, where the rotation $\bOmega = \Omega_Y \be_Y$ with its dependence on particle orientation $\psi = \arccos \lb( \bp \cdot \bgh \rb)$ shown in Fig.~\ref{fig:fig6a}.

We next find the velocity changes $\Delta \bU$ by evaluating the integral in \eqref{eqn:DeltaU_recip} in the outer region, as it is apparent by now that if $\bOmega$ is determined by outer region so is $\Delta \bU$. Hence in the outer region, we simplify the integral in \eqref{eqn:DeltaU_recip} like that mentioned in previous section, rescale the variables and transform the integral to Fourier space to derive
\begin{equation}
    \Delta \bU =  \frac{Ri_v^{2/3}}{8\pi^3} \bgh \cdot  \int{ \rhoh' \lb( \bk \rb) \frac{1}{k^2} \lb( \bI - \frac{\bk \bk}{k^2} \rb) d\bk }.
\end{equation}
Again the rescaling alone reveals the dependence of $\Delta \bU$ on $Ri_v$, with the remaining integral bringing in the dependence on $\beta$ and $\psi$. We evaluate this integral in the particle-aligned coordinate system with the leading order density $\rhoh'_{00}$. As the density transport is governed by advection in this limit, the leading order density ${\hat{\rho}}'_{00}$ lacks any mirror-symmetry and hence is sufficient to determine finite velocity changes. Numerically evaluating the integral yields $\Delta U_Y = 0$ but finite $\Delta U_X$, $\Delta U_Z$ that are linear in $\beta$ and fit well with $0.3485 Ri_v^{2/3} \beta \sin \psi \cos \psi$ and $0.3007 Ri_v^{2/3} \beta \left( \cos^2 \psi + 2/9 \right)$, respectively. See Figs.~\ref{fig:fig6b}, \ref{fig:fig6c}. Combining all three components, the velocity changes in this limit are
\begin{multline}
\Delta \bU \approx Ri_v^{2/3} \beta \Biggl\{ -0.3485 \left( \bp \cdot \hat\bg \right) \left( \left( \bp \cdot \hat\bg \right)  \bp -  {\hat\bg}\right) \\+ 0.3007 \left( \left( \bp \cdot \hat\bg \right)^2 + \frac{2}{9} \right) \bp \Biggr\}.
\label{eqn:DeltaU-LargePe}
\end{multline}
Analytical integration is only possible for vertical swimming and it yields the velocity
\begin{equation}
    \bU = \left\{ 1 + \frac{15 \sqrt{3}}{56 \pi} \Gamma\left( \frac{2}{3} \right)^3 Ri_v^{2/3} \beta \right\} \bp
    \label{eqn:DU-Vertical-largePe}
\end{equation}
that is in line with the previous work \cite{Shaik2021}. Speed changes in this configuration are again insensitive to whether the swimming is up or down the density gradients. Pullers swimming vertically simply speed up relative to their speed in the absence of density gradients. On the other hand, numerical integration for horizontal swimming yields the velocity
\begin{equation}
\bU = \left\{ 1 + 0.07060 \, Ri_v^{2/3} \beta \right\} \bp,
\end{equation}
ultimately unraveling the slow down of pushers swimming in this direction.

\bibliography{references}% Produces the bibliography via BibTeX.

\end{document}